\documentclass{rspublic}
\usepackage{bm}

\begin{document}

\title[One Gravitational Potential or Two?]{One Gravitational Potential or Two?  Forecasts and Tests}

\author[E. Bertschinger]{Edmund Bertschinger}

\affiliation{Department of Physics and Kavli Institute for Astrophysics and Space Research, Massachusetts Institute of Technology, Cambridge, MA 02139}

\label{firstpage}

\maketitle

\begin{abstract}{gravitation, cosmological tests, general relativity}
The metric of a perturbed Robertson-Walker spacetime is characterized by three functions: a scale-factor giving the expansion history and two potentials which generalize the single potential of Newtonian gravity.  The Newtonian potential induces peculiar velocities and, from these, the growth of matter fluctuations.  Massless particles respond equally to the Newtonian potential and to a curvature potential.  The difference of the two potentials, called the gravitational slip, is predicted to be very small in general relativity but can be substantial in modified gravity theories.  The two potentials can be measured, and gravity tested on cosmological scales, by combining weak gravitational lensing or the Integrated Sachs-Wolfe effect with galaxy peculiar velocities or clustering.
\end{abstract}

\section{Introduction}
Cosmic acceleration remains as mysterious today as it was when its discovery was announced in 1998 (Riess {\it et al.} 1998; Perlmutter {\it et al.} 1999).  In the cosmological standard model based on general relativity, a homogeneous, isotropic, and uniformly expanding universe accelerates if and only if the pressure is large and negative such that $\rho+3p<0$.  Dark energy is the name given to a class of exotic fluids having this property.  Dark energy may be either constant (the cosmological constant $\Lambda$) or have varying density and pressure (dynamical dark energy).  At present, the chief goal of observational cosmology is to measure the expansion history $H(a)$ in order to determine the time-dependence of the energy density $\rho(a)$ from the Friedmann equation
\begin{equation}\label{Friedmann}
  H^2(a)=\frac{8\pi}{3}G\rho(a)-Ka^{-2}\ .
\end{equation}
Measurements of $\rho(a)$ yield the dark energy equation of state $p(\rho)$ through energy conservation, $d\rho/d\ln a=-3(\rho+p)$.

The conclusion that dark energy exists depends on the assumptions of the cosmological standard model.  For example, if the galaxy distribution is strongly inhomogeneous while preserving near-isotropy around us, it is possible for galaxy motions along the past light cone to mimic the acceleration of a homogenous universe without dark energy (Kolb {\it et al.} 2005; Vanderveld {\it et al.} 2006; Clifton {\it et al.} 2008).  This possibility requires very special initial conditions that break the fundamental translation-invariance of the standard model; moreover, gravitational instability might convert large radial gradients into large angular gradients that are not seen.  The breaking of translation invariance would be problematic for initial conditions generated by inflation, among other challenges.  This article will therefore retain the assumption of large-scale homogeneity and isotropy present in the standard cosmological model.

There is one key element of the standard cosmological model that is poorly tested: general relativity (GR) itself.  Since Einstein's 1917 paper introducing the cosmological constant, GR has been so fundamental to cosmological models that it is not immediately obvious how to modify it.  Clearly any modifications great enough to eliminate the need for dark energy must preserve the successes of GR in the solar system and in binary pulsars (Will 2006).  Thus the modifications of the field equations or their solution must be scale-dependent: negligible on short length scales and appreciable on the Hubble scale.

Measurements of the expansion history alone cannot distinguish GR from alternative theories.  An example of a modified gravity theory having cosmic acceleration without dark energy is the DGP braneworld model of Dvali {\it et al.} (2000), implemented in cosmology by Deffayet {\it et al.} (2002).  This is a higher-dimensional theory in which our universe lives on a brane of three spatial dimensions.  The field equations are modified at long wavelengths.  In the so-called self-accelerating branch of the theory, the Friedmann equation is modified to become
\begin{equation}\label{DGPFriedmann}
  H(H-H_\infty)=\frac{8\pi}{3}G\rho(a)-Ka^{-2}\ ,
\end{equation}
where $H_\infty$ is a constant.  At early times $H>H_\infty$.  As the universe expands, $\rho(a)$ decreases so $H$ decreases until it asymptotically approaches $H_\infty$.  A constant Hubble parameter $H=d\ln a/dt$ implies exponential expansion, the same as a cosmological constant-dominated universe in GR.  Such a model has acceleration without dark energy.

The DGP model makes a specific prediction for $H(a)$ that can be tested and falsified; it appears to be in conflict with observations (Fairbairn \& Goobar 2006; Lombriser {\it et al.} 2009).  However, there are other classes of modified gravity theories that are not, and cannot be, ruled out based on the expansion history.  In particular, models in which the Einstein-Hilbert action for gravity is modified by an arbitrary function of the Ricci scalar $R$ can reproduce any $H(a)$ (Multamaki \& Vilja 2006; Capozziello {\it et al.} 2006).  The action for the theory is
\begin{equation}\label{f(R)action}
  S[g_{\mu\nu}(x)]=\int\frac{d^4x\,\sqrt{-g}}{16\pi G}[R+f(R)]
\end{equation}
where $f(R)=0$ for GR.  There is no fundamental theory for $f(R)$; instead, there exist parametrized forms (e.g. Hu \& Sawicki 2007a; Starobinsky 2007) for which predictions can be made and compared with observations.

These results imply that GR cannot be tested on cosmological scales using the expansion history.  One must look beyond the homogeneous and isotropic, uniformly expanding Robertson-Walker models to include perturbations.  Moreover, it is most helpful to identify those features of the data that most distinguish modified gravity theories from GR, and use them to construct cosmological tests.  Theorists and observers have begun to do this (e.g., Zhang {\it et al.} 2007; Reyes {\it et al.} 2010).  As we show, these tests involve two functions of space and time, which we call potentials.  We also show that observations of galaxies or cold dark matter, in particular the growth of structure, are insufficient to test GR.

What does it mean to test GR?  General relativity makes a number of assumptions that may also hold in alternative theories of gravity (Bertschinger 2006a):
\begin{enumerate}
  \item Spacetime is a 4-dimensional pseudo-Riemannian manifold.
  \item Special relativity holds locally.  In particular, energy-momentum is locally conserved.
  \item The weak equivalence principle holds, i.e., freely-falling bodies follow spacetime geodesics.
  \item The metric is the soution to the Einstein field equations subject to appropriate initial and boundary conditions; this is uniquely true in GR.
\end{enumerate}
The first three assumptions will be adopted here for modified gravity theories as well as for GR.  Modifying gravity then implies modifying the field equations; the fourth assumption is to be tested.  {\bf Specifically, testing GR requires measuring the metric (or the transfer functions and power spectra for perturbations in the metric) and comparing measurements with the solution of the field equations.}

Section 2  introduces the two potentials used to characterize cosmological perturbations and shows that nonrelativistic matter (atoms and cold dark matter) are sensitive to only one of them while the deflection and redshift of photons are sensitive to both potentials.  Section 3 introduces the primary tools for measuring the potentials: peculiar velocities, the growth of structure, the Integrated Sachs-Wolfe effect, and gravitational lensing.  While full reconstruction of the potentials remains a longterm goal, current efforts emphasize measurements of correlation functions and power spectra, which also provide valuable constraints on the two potentials.

Section 4 summarizes the field equations for the potentials in GR and in one class of modified gravity theories, the $f(R)$ theories.  It is shown that a key discriminating feature between theories is the ``gravitational slip,'' the difference of the two potentials.  In GR the slip is generated only by relativistic shear stress and is negligible in the matter-dominated era. In $f(R)$ theories the slip is generated by a new gravitational field called the scalaron.  The scalaron also modifies the growth of structure while avoiding solar-system constraints.   Section 5 briefly summarizes current tests.

More extensive reviews are given by Caldwell \& Kamionkowski (2009), Silvestri \& Trodden (2009), Uzan (2009), and Jain \& Khoury (2010).

\section{Two potentials}

The most general metric of a weakly perturbed Robertson-Walker spacetime --- regardless of whether GR is valid --- takes the form
\begin{equation}\label{pertRW1}
  ds^2=a^2(\tau)\left\{-(1+2\Phi)d\tau^2+2w_idx^id\tau+[(1-2\Psi)\gamma_{ij}+2s_{ij}]dx^idx^j
    \right\}
\end{equation}
where $\gamma_{ij}$ is the three-metric of constant-curvature spaces, e.g. $\gamma_{ij}=\delta_{ij}$ for Cartesian coordinates in a Euclidean space.  The inverse spatial metric is $\gamma^{ij}$.  For an unperturbed Robertson-Walker spacetime, $\Phi=\Psi=w_i=s_{ij}=0$.  Without loss of generality we can take $s_{ij}$ to be symmetric and traceless, i.e., $\gamma^{ij}s_{ij}=0$.  The time variable $\tau$ is called conformal time and is related to the proper time for comoving observers (those at fixed spatial coordinates) in the unperturbed background by $dt=a\,d\tau$.

Equation (\ref{pertRW1}) is too general in that the same spacetime is described by infinitely many sets of solutions for the functions $(\Phi,\Psi,w_i,s_{ij})$.  This ambiguity arises because of general covariance, i.e. the invariance of the gravitational action under coordinate transformations.  Like the vector potential $A_\mu$ of electromagnetism, the metric perturbations change under gauge transformations (here, infinitesimal coordinate transformations).  We are free to remove the ambiguity by gauge-fixing, which means choosing a particular coordinate system.  We choose the Poisson gauge (Bertschinger 1996), a coordinate system defined by transverse gauge conditions
\begin{equation}\label{Poissongauge}
  \nabla^iw_i=0\ ,\ \ \nabla^j s_{ij}=0\ ,
\end{equation}
where spatial divergences are covariant derivatives taken with respect to the background 3-metric $\gamma_{ij}$.  The function $w_i$ is called the gravomagnetic potential and is associated with vector perturbations.  The function $s_{ij}$ is called the gravitational wave strain and is associated with tensor perturbations.  The Poisson gauge generalizes the transverse-traceless gauge widely used for gravitational waves to include all possible perturbations of the metric (scalar, vector, and tensor).

Although gravomagnetic and gravitational wave perturbations are expected to be present in the universe, they have not been directly detected and are known to be much smaller in amplitude than the other terms in the metric.  Therefore we neglect them in what follows.

The potential $\Phi$ is called the Newtonian potential.  Being part of the time-time part of the metric, it is important for the motion of nonrelativistic particles.  For slowly moving particles, $dx^i/d\tau$ is small and the spatial part of the metric makes a negligible contribution to the proper time and therefore to the equations of motion that extremize proper time.  The second potential $\Psi$ is called the curvature potential because it is associated with spatial curvature.  It is important for the motion of relativistic particles.

With this coordinate choice and with the neglect of gravomagnetic and gravitational wave perturbations, the metric simplifies to the Conformal Newtonian metric (Mukhanov {\it et al.} 1992, Ma \& Bertschinger 1995)
\begin{equation}\label{cnmetric}
  ds^2=a^2(\tau)\left[-(1+2\Phi)d\tau^2+(1-2\Psi)\gamma_{ij}dx^idx^j\right]\ .
\end{equation}
The potentials are assumed to be small enough ($\Phi\sim\Psi\sim10^{-5}$) that their products can be neglected.  For a flat background, with $\gamma_{ij}=\delta_{ij}$, and with $\Psi=\Phi$, this metric is conformal to (i.e., identical to aside from an overall multiplicative factor) the weak-field metric used in the Newtonian limit (Hartle 2003).  The conformal Newtonian coordinate system used here has the advantage of making motion look simple.  The reader should beware that different choices of names and signs for the two potentials appear in the literature.

Freely-falling particles move along geodesics of metric (\ref{cnmetric}) and obey equations of motion
\begin{subequations}\label{geodesic}
\begin{eqnarray}
  \frac{dx^i}{d\tau}&=&(1+\Phi+\Psi)v^i\ ,\label{pecvel}\\
  \frac{1}{\gamma a(1-\Psi)}\frac{d}{d\tau}\left[\gamma a(1-\Psi) v^i\right]&=&
    -\nabla^i(\Phi+v^2\Psi)-(1+\Phi+\Psi)\gamma^i_{jk}v^jv^k\ .\quad\quad \label{pecaccel} 
\end{eqnarray}
\end{subequations}
Here, $v^i$ is the proper (physical) 3-velocity measured by a comoving observer (note that $\tau$ is not the proper time).  Specifically, it is the {\it peculiar velocity} with respect to uniform Hubble flow.  Other terms in the equations of motion have their usual definitions: $v^2\equiv\gamma_{ij}v^iv^j$, $\gamma\equiv(1-v^2)^{-1/2}$ is the special relativistic Lorentz factor, and $\gamma^i_{jk}$ is the spatial connection needed when non-Cartesian coordinates are used (giving rise to terms such as centripetal acceleration).  The factors $(1+\Phi+\Psi)$ convert proper velocity to coordinate velocity.  Hereafter we will ignore the $\sim10^{-5}$ corrections from proper to coordinate velocities.

The equations of motion simplify in the limits $v\to0$ (nonrelativistic particles) and $v\to1$ (massless particles).  In the first case, denoted cold dark matter (CDM),
\begin{equation}\label{cdmeom}
   \frac{1}{a}\frac{d(a{\bm v})}{d\tau}=-{\bm\nabla}\Phi\ ,\ \ v^2\ll 1\ \ \hbox{(CDM)}\ .
\end{equation}
In  the second case,
\begin{equation}\label{photeom}
  \frac{d{\bm v}}{d\tau}=-{\bm\nabla}_\perp(\Phi+\Psi)\ ,\ \ v^2=1\ \ \hbox{(photons)}
\end{equation}
where the gradient is taken in the plane perpendicular to the photon trajectory:
\begin{equation}\label{gradperp}
 \nabla^i_\perp\equiv\nabla^i-v^iv_j\nabla^j\ .
\end{equation}

Equations (\ref{cdmeom}) and (\ref{photeom}) are very important and are more general than GR.  They
have a crucial implication: {\bf The dynamics of galaxies and CDM cannot fully test GR because they depend on only one of the two potentials.}

Gravitational lensing measurements are sensitive to the sum of the two potentials.  Another dependence arises for the redshift of light:
\begin{equation}\label{gredshift}
  \frac{d\ln(aE)}{d\tau}=\frac{\partial\Psi}{\partial\tau}-{\bm v}\cdot{\bm\nabla}\Phi\ ,\ \ v^2=1\ \ \hbox{(photons)}\ .
\end{equation}
The first term on the right-hand side is less familiar than the second term (gravitational redshift).  It becomes more familiar when equation (\ref{gredshift}) is integrated to give the accumulated change in energy of a photon between emission at redshift $z$ and observation:
\begin{equation}\label{redshift}
  (1+z)\frac{E_{\rm obs}}{E_{\rm em}}=1+\Phi_{\rm  em}-\Phi_{\rm obs}
    -\int_0^z(\dot\Phi+\dot\Psi)(d\tau/dz)\,dz\ ,
\end{equation}
where a dot denotes $\partial/\partial\tau$.  Equation (\ref{redshift}) gives the familiar Sachs-Wolfe effect (Sachs \& Wolfe 1967); the last term, arising from time-variation of the potentials along the past light cone, is called the Integrated Sachs-Wolfe (ISW) effect.  Thus cosmic microwave background (CMB) anisotropy responds to both the Newtonian potential $\Phi_{\rm em}$ and to the time derivative of the sum of the two potentials.

\section{Methods to measure the potentials}

Testing gravity on cosmological scales requires making distinct measurements for each of the three functions appearing in the metric.  The most direct way is:
\begin{enumerate}
  \item Measure $a(\tau)$ using redshift and distance measurements to infer the cosmic expansion history $H(a)$; integrate $d\tau/da=1/(Ha^2)$ to get $\tau(a)$ which can then be inverted.
  \item Measure $\Phi({\bm x},\tau)$ using equation (\ref{cdmeom}) with the peculiar velocities of galaxies.
  \item Measure $(\Phi+\Psi)({\bm  x},\tau)$ using equation (\ref{photeom}) with gravitational lensing or equation (\ref{redshift}) with the ISW effect.
\end{enumerate}
Although the second two measurements give derivatives of the potentials rather than their absolute values, with appropriate initial or boundary conditions the potentials can be determined up to irrelevant constants.  These measurements are then compared with the relationships between the three functions predicted by GR or modified gravity theories.

In practice these direct methods are difficult at best.  Peculiar velocities are difficult to measure with sufficient precision for gravitational tests; while the radial velocity is easy to measure from redshifts, the Hubble velocity is proportional to distance.  Even with the precision given by Type Ia supernovae, uncertainties in distance measurements exceed the peculiar velocities for distances beyond a few tens of Mpc.  Gravitational lensing is promising, however the deflections are small for structures larger than a few Mpc.  The ISW signal is concentrated at the lowest multipoles of the CMB where primary anistotropy and cosmic variance (the variance arising from Gaussian random processes with a very small number of terms contributing to the signal) are large.

\subsection{Growth of structure}

To increase the power of galaxies to measure $\Phi$, cosmologists commonly turn to the growth of density perturbations. Density perturbations are part of the stress-energy tensor which is decomposed for scalar perturbations as
\begin{subequations}\label{emtensor}
\begin{eqnarray}
  T^0_{\ \,0}&=&-(\bar\rho+\delta\rho)\ ,\ \label{t00}\\
  T^0_{\ \,i}&=&-(\bar\rho+\bar p)\nabla_i u\ ,\label{t0i}\\
  T^i_{\ \,j}&=&\delta^i_{\ \,j}(\bar p+\delta p)+\frac{1}{2}
    (\bar\rho+\bar p)\left(\nabla^i\nabla_j-\frac{1}{3}\delta^i_{\ \,j}
     \Delta\right)\pi\label{tij}
\end{eqnarray}
\end{subequations}
where $\Delta=\gamma^{ij}\nabla_i\nabla_j$ is the spatial Laplacian, $\bar\rho$ and $\bar p$ are the background density and pressure and $u$ and $\pi$ are velocity and shear stress potentials, respectively.  The perturbations are gauge-dependent.  A convenient gauge-invariant density perturbation is the physical number density perturbation in the fluid rest frame (Bertschinger 2006b), obtained from conformal Newtonian gauge variables as
\begin{equation}\label{nudef}
  \nu\equiv\frac{\delta\rho+3{\cal H}u}{\bar\rho+\bar p}\ ,\ \ {\cal H}=\frac{\dot a}{a}=aH\ .
\end{equation}
For a nonrelativistic fluid on scales much less than the Hubble length with negligible shear stress, linear perturbations of energy-momentum conservation gives
\begin{equation}\label{denevol}
  \frac{1}{a}\frac{\partial}{\partial\tau}\left(a\frac{\partial\nu}{\partial\tau}\right)=\Delta(\Phi+
    c_s^2\nu+\sigma)\ .
\end{equation}
Here $c_s=(d\bar p/d\bar\rho)^{1/2}$ is the adiabatic sound speed and the dimensionless entropy perturbation is
\begin{equation}\label{entropy}
  \sigma=\frac{\delta p-c_s^2\delta\rho}{\bar\rho+\bar p}\ .
\end{equation}
Equation (\ref{denevol}) says that the density perturbation growth is driven by the Newtonian potential $\Phi$, it is opposed by adiabatic pressure perturbations through the Jeans term $c_s^2\nu$, and responds to pressure perturbations produced by entropy perturbations if they are present.  It is important to note that equation (\ref{denevol}) is restricted to linear density perturbations.  It may be generalized to incorporate nonlinear perturbations on scales much smaller than the Hubble length, as is commonly done in cosmological simulation codes (see Bertschinger 1998 for a review).

If pressure forces are unimportant on linear scales --- as is thought to be valid on scales much larger than 10 Mpc, then atomic gas, galaxies, and cold dark matter particles all approximately obey equation (\ref{denevol}) with $c_s^2=\sigma=0$.   Two key assumptions are implicit:
\begin{enumerate}
  \item All matter couples universally to gravity; there are no additional long-range forces.
  \item Luminous atoms trace dark matter and so do galaxies, or one can correct for ``biased galaxy formation'' (Dekel \& Rees 1987; Smith {\it et al.} 2007).
\end{enumerate}
The first assumption involves fundamental physics and is a restatement of the weak equivalence principle.  It is important to test this assumption on cosmological scales but that is beyond the scope of this article (Jain \& Khoury 2010).  The second assumption is astrophysical and represents a serious concern for cosmological tests.

If these assumptions can be justified, then growth of structure offers an alternative means to determining the Newtonian potential $\Phi$.  A measurement of density perturbations $\nu(a)$ substituted into equation (\ref{denevol}) gives $\Delta\Phi$.

\subsection{Weak gravitational lensing}

Weak gravitational lensing results when the displacements of photon trajectories implied by equation (\ref{photeom}) accumulate over the past light cone so as to deform the images of distant sources.  The angular deflection arising between $z_L$ and $z_L+\delta z_L$ is
\begin{equation}\label{deflection}
  \delta{\bm\alpha}=-{\bm\nabla}_\perp(\Phi+\Psi)\frac{d\tau}{dz_L}\delta z_L
\end{equation}
where ${\bm\alpha}$ is a two-dimensional vector perpendicular to the photon trajectory.  The gravitational lens equation (Bartelmann \& Schneider 2001)
\begin{equation}\label{lenseq}
  {\bm\beta}={\bm\theta}-\frac{r_{LS}}{r_S}\delta{\bm\alpha}
\end{equation}
relates the observed direction of a ray ${\bm\theta}$ to the deflection and the direction to the source ${\bm\beta}$ in the absence of deflection.  The two distances $r_S$ and $r_{LS}$ are the comoving angular distances $r(\chi)$ appearing in the spatial line element $\gamma_{ij}dx^idx^j=d\chi^2+r^2(\chi)(d\theta^2+\sin^2\theta d\phi^2)$.

The absolute deflection of rays is unobservable since we have no way of knowing the true direction of sources in the absence of deflection. However, the gradient of deflection results in image magnification and shear that are observable and are described by the inverse magnification matrix
\begin{equation}\label{magnify}
  {\sf M}^{-1}=\frac{\partial{\bm\beta}}{\partial{\bm\theta}}={\bm 1}+
    \int_0^{z_S}\frac{r_Lr_{LS}}{r_S}{\bm\nabla}_\perp{\bm\nabla}_\perp(\Phi+\Psi)
    \frac{d\tau}{dz_L}dz_L\ .
\end{equation}
The two-by-two magnification matrix is parametrized by a convergence $\kappa$ and shear parameters $\gamma_1$ and $\gamma_2$:
\begin{equation}\label{shearmatrix}
  {\sf M}=\left(
    \begin{array}{cc}
      1+\kappa+\gamma_1 & \gamma_2 \\
      \gamma_2 & 1+\kappa-\gamma_1
    \end{array}
    \right)\ .
\end{equation}
This description is most useful in the weak lensing limit $\kappa^2\ll1$, $\gamma_i^2\ll1$.  Although the convergence $\kappa$ cannot be measured directly because the unlensed size and flux of sources is generally unknown, the shear components $\gamma_i$ can be measured by averaging galaxy image distortions over fields of view containing hundreds of distant galaxies (Hoekstra \& Jain 2008).  Thus weak lensing offers a probe of the two-by-two traceless Hessian matrix of the sum of the two potentials:
\begin{equation}\label{potderivs}
  \Psi_{ij}=\left(\nabla_i\nabla_j-\frac{1}{2}\delta_{ij}\Delta_2\right)(\Phi+\Psi)
\end{equation}
where $\Delta_2={\bm\nabla}_\perp\cdot{\bm\nabla}_\perp$.  Given this matrix as a function of direction, and imposing appropriate boundary conditions, it is possible to invert the partial derivative operator to solve for $\Phi+\Psi$ or the convergence $\kappa$ (Kaiser \& Squires 1993).

\subsection{Power spectra and transfer functions}

An ideal test of gravitation theories would determine $\Phi({\bm x},\tau)$ and $\Psi({\bm x},\tau)$ along the past light cone giving (with the expansion history) the solution to the gravitational field equations over as much of spacetime as is accessible to our view.  Indeed, reconstruction should be the goal.  However, many problems impede this effort: galaxies (the main cosmological probe) are sparsely distributed through space, their properties and small-scale clustering are strongly affected by nonlinear dynamics and the astrophysics of star formation (e.g., Smith {\it et al.} 2009), and building up complete samples is an ongoing project for the field of observational cosmology.  In the case of CMB anisotropy, accurate measurements are possible but the Sachs-Wolfe and ISW effects cannot be separated from other sources of anisotropy generated during recombination and before.  Even if they could be separated, the ISW effect is most important at low multipoles where cosmological theories predict that the power spectrum has a chi-squared distribution with $2l+1$ degrees of freedom (cosmic variance).

For these reasons, it is useful to measure the statistical properties of the galaxy density and peculiar velocity fields, weak lensing shear, and the ISW effect by measuring correlation functions.  Correlation functions can be regarded as a lossy data compression technique; they eliminate information about non-Gaussianity but facilitate tests of the time- and length-scale dependence of fluctuations and thereby of the metric potentials.  Each fluctuating field ($\Phi$, $\Psi$, $\nu$, $u$, $\gamma_i$, etc.) has an autocorrelation with itself and cross-correlations with other fields.  The Fourier transform of a spatial correlation function is a power spectrum or cross-power spectrum, which itself is related to a correlation in Fourier space.  For a Euclidean background,
\begin{equation}\label{correl}
  \xi_{\nu\nu}(\vert{\bm x}_1-{\bm x}_2\vert)
  \equiv \left\langle\nu({\bm x}_1)\nu({\bm x}_2)\right\rangle
 =\int\frac{d^3k}{(2\pi)^3}\,e^{i{\bm k}\cdot({\bm x}_1-{\bm x}_2)}\,P_{\nu\nu}(k)\ ,
\end{equation}
where
\begin{equation}\label{powerspec}
  \left\langle\nu({\bm k})\nu({\bm k}')\right\rangle=(2\pi)^3P_{\nu\nu}(k)\delta^3({\bm k}+{\bm k}')\ .
\end{equation}
These equations assume that the fluctuations are statistically homogeneous and isotropic, which holds for quantum fluctuations superposed on a translationally- and rotationally-invariant background.

The time-dependence of the fields and correlations has been suppressed in equations (\ref{correl})--(\ref{powerspec}).  In linear perturbation theory, all fluctuation fields are given by convolutions of a primordial scalar field or possibly a superposition of several such convolutions.  The simplest category is curvature fluctuations characterized by the gauge-invariant spatial curvature perturbation
\begin{equation}\label{Rcurv}
  {\cal R}=\Psi+{\cal H}u\ ,
\end{equation}
where $u$ is the total velocity potential (a weighted average over all species) defined by equation (\ref{t0i}).  The field ${\cal R}$ is constant on scales larger than the Hubble and curvature lengths irrespective of the expansion history (Lyth 1985).  It is even constant on super-horizon scales in modified gravity theories (Bertschinger 2006a).

Every other field is given, in Fourier space, by a rotationally invariant transfer function multiplying this primary variable:
\begin{equation}\label{transfer}
  \nu({\bm k},\tau)=\nu(k,\tau){\cal R}({\bm k})\ .
\end{equation}
The transfer functions for all matter and metric variables are given by solutions of the appropriate field equations with initial conditions ${\cal R}({\bm k})=1$.  The power spectrum of ${\cal R}$ is conventionally made nondimensional by defining
\begin{equation}\label{DeltaR}
  \Delta_{\cal R}^2(k)=\frac{4\pi k^3P_{\cal RR}(k)}{(2\pi)^3}\ .
\end{equation}
WMAP 7-year results (Larson {\it et al.} 2011) give $\Delta_{\cal R}=(4.93\pm 0.11)\times 10^{-5}$ for $k=0.002$ Mpc$^{-1}$.

Cross-correlation techniques can be used to extract information about the ISW effect that is otherwise hidden by other sources of low-multipole CMB anisotropy.  Galaxy (mass) density fluctuations at modest redshift correlate with the potential fluctuations giving rise to the ISW effect.  The cross-correlation of CMB anisotropy with galaxy density can therefore extract the ISW signal (Crittenden \& Turok 1996; Afshordi {\it et al.} 2004).

Cross-correlation techniques can also be used to minimize the systematic effects of galaxy bias.  Zhang {\it et al.} (2007) defined a ratio of weak lensing shear to galaxy peculiar velocities based on cross-correlations with the galaxy density.  The resulting statistic, $E_G$, can be measured on large scales using weak lensing combined with redshift-space distortions in galaxy redshift surveys.

\section{Potentials in GR and in $f(R)$ theories}

To test a theory of gravity, test the gravitational field equations.  As argued in the Introduction, the field equations for the unperturbed background are insufficient.  Therefore we examine the linear perturbations of the field equations.  In general relativity the field equations are $G^\mu_{\ \,\nu}=8\pi GT^\mu_{\ \,\nu}$, with first-order perturbations giving
\begin{subequations}\label{eins}
\begin{eqnarray}
  (\Delta+3K)\Psi&=&\alpha\nu\ ,\ \label{g00}\quad\\
  \dot\Psi+{\cal H}\Phi&=&\alpha u\ ,\ \label{g0i}\\
  \Psi-\Phi&=&\alpha\pi\ ,\label{gij}\\
  \ddot\Psi-K\Psi+{\cal H}(2\dot\Psi+\dot\Phi)\!\!\!\!&+&\!\!\!\!\left(2\dot{\cal H}
   +{\cal H}^2\right)\Phi+\frac{1}{3}\Delta(\Phi-\Psi)=\frac{\alpha\delta p}
    {\bar\rho+\bar p}\ ,\label{gkk}
\end{eqnarray}
\end{subequations}
where
\begin{equation}\label{alphadef}
  \alpha\equiv4\pi Ga^2(\bar\rho+\bar p)\ .
\end{equation}
The right-hand side of equations (\ref{eins}) uses the stress-energy perturbations given in equations (\ref{emtensor}) and (\ref{nudef}).  An alternative form of equation (\ref{gkk}) is (Bertschinger 2006a)
\begin{equation}\label{psieom}
  \frac{\alpha}{\cal H}\frac{\partial}{\partial\tau}\left[\frac{{\cal H}^2}{\alpha a^2}
    \frac{\partial}{\partial\tau}\left(\frac{a^2}{\cal H}\Psi\right)\right]-c_s^2\Delta\Psi
   =\alpha S\ ,\ \  S\equiv\sigma+\frac{1}{\cal H}\frac{\partial}{\partial\tau}
    \left({\cal H}^2\pi\right)+\frac{1}{3}\Delta\pi\ .
\end{equation}
This equation is exact in general relativity with linear perturbation theory for open, curved, or flat backgrounds, and for arbitrary matter fields and dynamics.

If we know nothing about the mass-energy content of the universe, then the Einstein equations provide no possible test; we cannot test GR without making assumptions about the mass-energy content.  We can simply measure the potentials and claim there exist matter fields obeying the Einstein field equations.  Because of the Bianchi identifies, we are even guaranteed that the ``stress-energy" tensor inferred this way is locally conserved.

This approach is unreasonable.  We aspire to a complete physical characterization of mass-energy in the laboratory as well as the cosmos.  Although we do not yet know the composition of the dark matter and dark energy, the most natural models share an important property: the shear stress is small compared with the energy density.  For example, for linear perturbations of a scalar field or a perfect fluid, $\pi=0$.  For free-streaming particles, $\Delta\pi\sim c_s^2\nu$ where $c_s$ is the  sound speed.  The largest source of shear stress in the standard cosmological model is relativistic neutrinos after neutrino decoupling.  On scales larger than the Hubble length, and while the neutrinos are relativistic (Ma \& Bertschinger 1995, with some change of notation),
\begin{equation}\label{nushear}
  \Delta\pi=\frac{2}{5}\frac{(\bar\rho+\bar p)_\nu}{(\bar\rho+\bar p)_{\rm tot}}\nu
\end{equation}
leading to (Bertschinger \& Zukin 2008)
\begin{equation}\label{nuslip}
  \frac{\Psi-\Phi}{\Phi}=\frac{2}{5}\frac{(\bar\rho+\bar p)_\nu}{(\bar\rho+\bar p)_{\rm tot}}\ .
\end{equation}
This ratio is negligible in the matter-dominated era. In order for shear stress to be gravitationally important, there must exist a relativistic component which dominates the mass-energy density of the universe  and which has nearly maximal free-streaming.  Suppose that it is the dark energy.  Were such a field present, $S$ in equation (\ref{psieom}) would be substantial and would modify the evolution of the gravitational potentials, and thereby would change the evolution of cold dark matter perturbations.  The observed structure formation could be achieved only if the shear stress becomes important after galaxies form (e.g., stressed dark energy, Calabrese {\it et al.} 2010).  Even in this case the ISW effect would be large.

The combination $\Psi-\Phi$ is called {\it gravitational slip} (Daniel {\it et al.} 2008; Daniel {\it et al.} 2009).  General relativity with the standard cosmological parameters predicts negligible gravitational slip today.

In alternative theories of gravity the slip is nonzero.  For example, in $f(R)$ theories (Sotiriou \& Faraoni 2010; De Felice \& Tsujikawa 2010), an additional scalar degree of freedom is present in the gravitational sector, called the scalaron $\chi$ (Starobinsky 1980).  In equation (\ref{f(R)action}), the modification is the term $f(R)$; the scalaron $\chi$ is given by the fluctuations in its derivative,
\begin{equation}\label{fRdef}
  f_R\equiv\frac{df}{dR}\ ,\ \ \chi\equiv\frac{df_R}{dR}\delta R\ .
\end{equation}
The scalaron obeys the wave equation of a field with mass
\begin{equation}\label{scalaronmass}
  m_s^2=\frac{1+f_R}{3df_R/dR}\ .
\end{equation}
The field equations (\ref{eins}) are modified by the addition of terms proportional to $\chi$ and derivatives of $f_R$.  For the present purposes the most important modification is to the gravitational slip:
\begin{equation}\label{gslip}
  (1+f_R)(\Psi-\Phi)=\chi+\alpha\pi\ .
\end{equation}
Modified gravity theories, and $f(R)$ in particular, modify the gravitational coupling and they introduce a nonzero gravitational slip independently of the matter sector.

How large is the gravitational slip predicted to be if $\pi=0$?  On scales smaller than the Hubble length, a quasi-static solution to the $f(R)$ field equations exists with (Hu \& Sawicki 2007b; Tsujikawa 2007; Pogosian \& Silvestri 2008)
\begin{equation}\label{f(R)slip}
  \chi\approx-\frac{(1+f_R)k^2}{k^2+a^2m_s^2}\frac{(\Phi+\Psi)}{3}\ .
\end{equation}
In addition, the Poisson equation (\ref{g00}) is modified on scales much smaller than the Hubble length to become
\begin{equation}\label{f(R)Poisson}
  \Delta\left(\frac{\Phi+\Psi}{2}\right)=\frac{\alpha\nu}{1+f_R}\ .
\end{equation}
Combining equations (\ref{gslip})--(\ref{f(R)slip}) gives
\begin{equation}\label{potslip}
  \frac{\Psi}{\Phi}=\frac{2k^2+3a^2m_s^2}{4k^2+3a^2m_s^2}\ .
\end{equation}
On such scales not only is the gravitational slip nonzero, but the effective gravitational coupling in the standard Poisson equation $\Delta\Phi=4\pi G_{\rm  eff}a^2\delta\rho$ is increased to
\begin{equation}\label{Geff}
  G_{\rm eff}=\frac{G}{1+f_R}\left(\frac{4k^2+3a^2m_s^2}{3k^2+3a^2m_s^2}\right)\ .
\end{equation}
On scales much larger than the scalaron Compton wavelength $m_s^{-1}$, gravity is unmodified aside from an overall reduction factor $1+f_R$.  However, on smaller scales the gravitational coupling increases by a factor $4/3$ and the gravitational slip becomes $\Psi-\Phi=-\frac{1}{2}\Phi$.

This is a substantial change in the behavior of gravity on small scales, suggesting that the $f(R)$ theories might be ruled out based on solar system tests.  However, the scalaron mass and the coupling factor $1+f_R$ depend on curvature and hence on local density.  The nonlinearity of the field equations allow for a gravitational ``chameleon'' effect (Khoury \& Weltman 2004) that shields gravity in the solar system (Hu \& Sawicki 2007a) by restricting the rapid change of $f_R$ to regions of high curvature.  It remains to be seen whether star formation and time-dependent stellar evolution is compatible with such modifications of gravity in stars.

\section{Current status of tests}

Strong limits on the gravitational slip are obtained in the solar system using the Shapiro time delay (Shapiro 1964), an effect in the propagation of light that is sensitive to $\Phi+\Psi$.  The round-trip time of signals from the Cassini spacecraft near Saturn as the signals pass close to the sun is longer than one would expect in flat spacetime.  The result is an extremely tight limit on the slip (Bertotti {\it et al.} 2003):
\begin{equation}\label{Cassini}
  \frac{\vert\Psi-\Phi\vert}{\Phi}<2\times10^{-5}\ \ \hbox{within the solar system}\ .
\end{equation}

On galaxy scales, Schwab {\it et al.} (2010) have combined stellar dynamics with strong gravitational lensing (e.g., the production of Einstein rings) for lenses in the Sloan Lens ACS Survey to limit the slip on kiloparsec scales:
\begin{equation}\label{SLACS}
  \frac{\Psi-\Phi}{\Phi}=0.01\pm0.05\ \ \hbox{within galaxies on kpc scales}\ .
\end{equation}

On scales of a few to a few tens of Mpc, cosmologists have recently been combining structure formation measurements with weak lensing  (Reyes {\it et al.} 2010; Daniel {\it et al.} 2010; Song {\it et al.} 2010) and lensing plus the ISW effect (Bean \& Tangmatitham 2010; Zhao {\it et al.} 2010; Daniel \& Linder 2010).  The Reyes {\it et al.} measurement was based on the $E_G$ parameter of Zhang {\it et al.} (2007), which has the virtue of relative insensitivity to galaxy bias and other systematics.  Ther results of these tests show consistency to date with GR on length scales up to about 50 Mpc.  Future surveys such as LSST (Abell {\it et al.} 2009) and Euclid  (Martinelli {\it et al.} 2011) will be able to improve the limits substantially.

\section {Acknowledgements}
The author thanks Alessandra Silvestri for her collaboration and helpful discussion.  This work was supported by NSF grant AST-0708501 and the Kavli Institute for Astrophysics and Space Research.

\section{References}
\begin{thedemobiblio}\smallskip{}

\item Abell, P. A. {\it et al.} 2009
	LSST Science Book, Version 2.0.
	arXiv:0912.0201.

\item Afshordi, N., Loh, Y.-S., \& Strauss, M. A. 2004
	Cross-correlation of the cosmic microwave background with the 2MASS galaxy survey:
	Signatures  of dark energy, hot gas, and point sources.
	{\it Phys. Rev.} D\ {\bf 69}, 083524.

\item Bartelmann, M., \& Schneider, P. 2001
	Weak gravitational lensing.
	{\it Phys. Rept.} {\bf 340}, 291-472.

\item Bean, R., \& Tangmatitham, M. 2010
	Current constraints on the cosmic growth history.
	{\it Phys. Rev.} D\ {\bf 81}, 083534.

\item Bertotti, B., Iess, L., \& Tortora, P. 2003
	A test of general relativity using radio links with the Cassini spacecraft.
	{\it Nature} {\bf 425}, 374-376.

\item Bertschinger, E. 1996
	Cosmological Dynamics.
	in {\it Cosmology and Large Scale Structure, proc. Les Houches Summer School, Session LX}
	ed. R. Schaeffer, J. Silk, M. Spiro, and J. Zinn-Justin (Amsterdam: Elsevier Science), 273-347.

\item Bertschinger, E. 1998
	Simulations of Structure Formation in the Universe.
	{\it Ann. Rev. Astron. Astrophys.} {\bf 36}, 599-654.

\item Bertschinger, E. 2006a
	On the Growth of Perturbations as a Test of Dark Energy and Gravity.
	{\it Ap. J.} {\bf 648}, 797-806.

\item Bertschinger, E. 2006b
	Effects of Cold Dark Matter Decoupling and Pair Annihilation on Cosmological Perturbations.
	{\it Phys. Rev.} D\ {\bf 74}, 063509.

\item Bertschinger, E., \& Zukin, P. 2008
	Distinguishing modified gravity from dark energy.
	{\it Phys. Rev.} D\ {\bf 78}, 024015.

\item Calabrese, E., de Putter, R., Huterer, D., Linder, E. V., \& Melchiorri, A. 2011
	Future CMB constraints on early, cold, or stressed dark energy.
	{\it Phys. Rev.} D\ {\bf 83}, 023011.

\item Caldwell, R. R., \& Kamionkowski, M. 2009
	The Physics of Cosmic Acceleration.
	{\it Ann. Rev. Nucl. Part. Sci.} {\bf 59}, 397-429.

\item Capozziello, S., Nojiri, S., Odintsov, S. D., \& Troisi, A. 2006
	Cosmological viability of f(R)-gravity as an ideal fluid and its compatibility with a
	matter dominated phase.
	{\it  Phys. Lett.} B\ {\bf 639}, 135-143.

\item Clifton, T., Ferreira, P. G., \& Land, K. 2008
	Living in a Void: Testing the Copernican Principle with Distant Supernovae.
	{\it Phys. Rev. Lett.} {\bf 101}, 131302.

\item Crittenden, R. G., \& Turok, N. 1996
	Looking for a Cosmological Constant with the Rees-Sciama Effect.
	{\it Phys. Rev. Lett.} {\bf 76}, 575-578.

\item Daniel, S. F., Caldwell, R. R., Cooray, A., \& Melchiorri, A. 2008
	Large scale structure as a probe of gravitational slip.
	{\it Phys. Rev.} D\ {\bf 77}, 103513.

\item Daniel, S. F., Caldwell, R. R., Cooray,  A., Serra, P., \& Melchiorri, A. 2009
	Multiparameter investigation of gravitational slip.
	{\it Phys. Rev.} D\ {\bf 80}, 023532.

\item Daniel, S. F., \&  Linder, E. V. 2010
	Confronting general relativity with further cosmological data.
	{\it Phys. Rev.} D\ {\bf  82}, 103523.

\item Daniel, S.F., Linder, E. V., Smith, T. L., Caldwell, R. R., Cooray,  A., Leauthaud, A.,
	\&  Lombriser, L. 2010
	Testing general relativity with current cosmological data.
	{\it Phys. Rev.}  D\ {\bf 81}, 123508.

\item De Felice, A., \& Tsujikawa, S. 2010
	f(R) Theories.
	{\it Living Rev. Relativ.} {\bf 13}, 3.

\item Deffayet, C., Dvali, G., \&  Gabadadze, G. 2002
	Accelerated universe from gravity leaking to extra dimensions.
	{\it Phys. Rev.} D\ {\bf 65}, 044023.

\item Dekel,  A., \& Rees, M. J. 1987
	Physical mechanisms for biased galaxy formation.
	{\it Nature}, {\bf 326}, 455-462.

\item Dvali, G., Gabadadze, G., \&  Porrati, M. 2000
	4D gravity on a brane in 5D Minkowski space.
	{\it Phys. Lett.} B\  {\bf 485}, 208-214.

\item Fairbairn, M., \& Goobar, A. 2006
	Supernova limits on brane world cosmology.
	{\it Phys. Lett.} B\ {\bf 642}, 432-435.

\item Hartle, J. B. 2003 {\it Gravity: An Introduction to Einstein's General Relativity}
	(San Francisco: Addison-Wesley).

\item Hoekstra, H., \& Jain, B. 2008
	Weak Gravitational Lensing and Its Cosmological Applications.
	{\it Ann. Rev. Nucl. Part. Sci.} {\bf 58}, 99-123.

\item Hu, W., \&  Sawicki, I. 2007a
	Models of f(R) cosmic acceleration that evade solar system tests.
	{\it Phys. Rev.} D\ {\bf 76}, 064004.

\item	Hu, W., \&  Sawicki, I. 2007b
	Parametrized post-Friedmann framework for modified gravity.
	{\it Phys. Rev.} D\ {\bf 76}, 104043.

\item	Jain, B., \& Khoury, J. 2010
	Cosmological tests of gravity.
	{\it Ann. Phys.} {\bf 325}, 1479-1516.

\item Kaiser, N., \& Squires, G. 1993
	Mapping the dark matter with weak gravitational lensing.
	{\it  Ap. J.} {\bf 404}, 441-450.

\item Khoury, J., \& Weltman, A. 2004
	Chameleon cosmology.
	{\it Phys. Rev.} D\ {\bf 69}, 044026.

\item Kolb, E. A., Matarrese, S., Notari, A., \& Riotto, A. 2005
	Effect of inhomogeneities on the expansion rate of the universe.
	{\it Phys. Rev.} D\ {\bf 71}, 023524.

\item Larson, D., Dunkley, J., Hinshaw, G., Komatsu, E., Nolta, M. R., Bennett, C. L., Gold, B.,
	Halpern, M., Hill, R. S., Jarosik, N., Kogut, A., Limon, M., Meyer, S. S., Odegard, N., Page, L.,
	Smith, K. M., Spergel, D. N., Tucker, G. S., Weiland, J. L., Wollack, W., \& Wright, E. L. 2011
	Seven-year Wilkinson Microwave Anisotropy Probe (WMAP) Observations: Power Spectra
	and WMAP-derived Parameters.
	{\it Ap. J. Suppl.} {\bf 192}, 16.

\item Lombriser, L., Hu, W., Fang, W., \& Seljak, U. 2009
	Cosmological constraints on DGP braneworld gravity with brane tension.
	{\it Phys. Rev.} D\ {\bf 80}, 063536.

\item Lyth, D. H. 1985
	Large-scale energy-density perturbations and inflation.
	{\it Phys. Rev.} D\ {\bf 31}, 1792-1798.

\item Ma, C.-P., \& Bertschinger, E. 1995
	Cosmological Perturbation Theory in the Synchronous and Conformal Newtonian Gauges.
	{\it Ap. J.} {\bf 455}, 7-25.

\item Martinelli, M., Calabrese, E., De Bernardis, F., Melchiorri,  A., Pagano, L., \& Scaramella, R. 2011
	Constraining modified gravitational theories with Euclid.
	{\it Phys. Rev.} D\ {\bf 83}, 023012.

\item Mukhanov, V. F., Feldman, H. A., \& Brandenberger, R. H. 1992
	Theory of cosmological perturbations.
	{\it Phys. Rept.} {\bf 215}, 203-333.

\item Multamaki, T., \& Vilja, I. 2006
	Cosmological expansion and the uniqueness of the gravitational action.
	{\it Phys. Rev.} D\ {\bf 73}, 024018.

\item Perlmutter, S., Aldering, G., Goldhaber, G., Knop, R. A., Nugent, P., Castro, P. G., Deustua, S.,
	Fabbro, S., Goobar, A., Groom, D. E., \& 23 coauthors 1999
	Measurements of Omega and Lambda from 42 High-Redshift Supernovae.
	{\it Ap. J.} {\bf 517}, 565-586.

\item Pogosian, L., \& Silvestri, A. 2008
	Pattern of growth in viable f(R) cosmologies.
	{\it Phys. Rev.} D\ {\bf 77}, 023503.

\item Reyes, R., Mandelbaum, R., Seljak, U., Baldauf, T., Gunn, J. E., Lombriser, L., \& Smith, R. E. 2010
	Confirmation of general relativity on large scales from weak lensing and galaxy velocities.
	{\it Nature}, {\bf 464}, 256-258.

\item Riess, A. G., Filippenko, A. V., Challis, P., Clocchiatti, A., Diercks, A., Garnavich, P. M.,
	Gilliland, R. L., Hogan, C. J., Jha, S., Kirshner, R. P., \& 10 coauthors 1998
	Observational Evidence from Supernovae for an Accelerating Universe and a
	Cosmological Constant.
	{\it Astron. J.} {\bf 116}, 1009-1038.

\item Sachs, R. K., \& Wolfe, A. M.  1967
	Perturbations of a Cosmological Model and Angular Variations of the Microwave Background.
	{\it Ap. J.} {\bf 147}, 73-90.

\item Schwab, J., Bolton, A. S., \& Rappaport, S. A. 2010
	Galaxy-Scale Strong-Lensing Tests of Gravity and Geometric Cosmology: Constraints and
	Systematic Limitations.
	{\it Ap. J.} {\bf 708}, 750-757.

\item Shapiro, I. 1964
	Fourth Test of General Relativity.
	{\it Phys. Rev. Lett.} {\bf 13}, 789-791.

\item Silvestri, A., \& Trodden, M. 2009
	Approaches to understanding cosmic acceleration.
	{\it Rept. Prog. Phys.} {\bf 72}, 096901.

\item Smith, R. E., Hernandez-Monteagudo, C., \& Seljak, U. 2009
	Impact of scale dependent bias and nonlinear structure growth on the integrated Sachs-Wolfe
	effect: Angular power spectra.
	{\it Phys. Rev.}  D\ {\bf 80}, 063528.

\item Smith, R. E., Scoccimarro, R., \& Sheth, R. K. 2007
	Scale dependence of halo and galaxy bias: Effects in real space.
	{\it  Phys. Rev.} D\ {\bf 75}, 063512.

\item Song, Y.-S., Zhao, G.-B., Bacon, D., Koyama, K., Nichol, R. C., \& Pogosian, L. 2010
	Complementarity of Weak Lensing and Peculiar Velocity Measurements in Testing General Relativity.
	arXiv:1011.2106.

\item Sotiriou, T. P., \& Faraoni, V. 2010
	f(R) theories of gravity.
	{\it Rev. Mod. Phys.} {\bf 82}, 451-497.

\item Starobinsky, A. A. 1980
	A new type of isotropic cosmological models without singularity.
	{\it Phys. Lett.} B\ {\bf 91}, 99-102.

\item Starobinsky, A. A. 2007
	Disappearing cosmological constant in f( R) gravity.
	{\it JETP Lett.}, {\bf 86}, 157-163.

\item Tsujikawa, S. 2007
	Matter density perturbations and effective gravitational constant in modified gravity models of
	dark energy.
	{\it Phys. Rev.} D\ {\bf 76}, 023514.

\item Uzan, J.-P. 2009
	Tests of General Relativity on Astrophysical Scales.
	{\it Gen. Rel. Grav.} {\it 42}, 2219-2246.

\item Vanderveld, R. A., Flanagan, E. E., \& Wasserman, I. 2006
	Mimicking dark energy with Lemaître-Tolman-Bondi models: Weak central singularities and
	critical points.
	{\it Phys. Rev.} D\ {\bf 74}, 023506.

\item Will, C. M. 2006
	The Confrontation between General Relativity and Experiment.
	{\it Living Rev. Relativ.} {\bf 9}, 3.

\item Zhang, P., Liguori, M., Bean, R., \& Dodelson, S. 2007
	Probing Gravity at Cosmological Scales by Measurements which Test the Relationship between
	Gravitational Lensing and Matter Overdensity.
	{\it Phys. Rev. Lett.} {\bf 99}, 141302.

\item Zhao, G.-B., Giannantonio, T., Pogosian, L., Silvestri, A., Bacon, D. J., Koyama, K., Nichol, R. C.,
	\& Song, Y.-S. 2010
	Probing modifications of general relativity using current cosmological observations.
	{\it Phys. Rev.} D\ {\bf 81}, 103510.
\end{thedemobiblio}

\end{document}